\documentclass[11pt]{article}
\usepackage{latexsym}
\usepackage[dvips]{graphics,color}
\usepackage{amsfonts}
\usepackage{amsmath,amssymb}
\usepackage{mathtools}
\usepackage{hyperref}
\hypersetup{unicode,colorlinks=true,linkcolor=red}
\newtheorem{theorem}{Theorem}
\newtheorem{proposition}[theorem]{Proposition}
\newtheorem{definition}{Definition}%
\newcommand{\beqr}{\begin{eqnarray}}
\newcommand{\feqr}{\end{eqnarray}}
\def\non{\nonumber}

\newcommand{\rf}[1]{(\ref{#1})}
\def\pr{^{\prime}}
\DeclareFontFamily{U}{eufm}{}
\DeclareFontShape{U}{eufm}{m}{n}{<->eufm10}{}
\DeclareSymbolFont{mcy}{U}{eufm}{m}{n}
\DeclareMathSymbol{\Hr}{\mathord}{mcy}{"58}

\def\pr#1#2#3{Phys. Rep. {\bf{#1}}, (#2) #3}

\def\prl#1#2#3{Phys. Rev. Lett. {\bf{#1}}, (#2) #3}

\def\ajp#1#2#3{Am. J. Phys. {\bf{#1}}, (#2) #3}
\def\ajm#1#2#3{Amer. J. Math. {\bf{#1}}, (#2) #3}

\def\pra#1#2#3{Phys. Rev. {A\bf{#1}}, (#2) #3}

\def\jmp#1#2#3{J. Math. Phys. {\bf{#1}}, (#2) #3}

\def\remp#1#2#3{Rep. Math. Phys. {\bf{#1}}, (#2) #3}

\def\jfaa#1#2#3{J. Fourier Anal. Appl. {A\bf{#1}}, (#2) #3}

\usepackage[active]{srcltx}
\begin{document}

\begin{center}
{\Large \bf The Non-commutative Robertson-Schr\"{o}dinger Uncertainty Principle}\\
[4mm]
{\large Agapitos N. Hatzinikitas} \\ [5mm]
{\small Department of Mathematics, \\ 
University of Aegean, \\
School of Sciences, \\
Karlovasi, 83200\\
Samos Greece \\
E-mail: ahatz@aegean.gr}\\ [5mm]
\end{center}


\abstract{We investigate properties of the covariance matrix in the framework of non-commutative quantum mechanics for an one-parameter family of transformations between the familiar Heisenberg-Weyl algebra and a particular extension of it. Employing as a measure of the Robertson-Schr\"{o}dinger uncertainty principle the linear symplectic capacity of the Weyl ellipsoid (and its dual), we determine its corresponding bounds. Inequalities between the capacities for non-commutative phase-spaces are established. We also present a constructive example based on a simple model to justify our theoretical predictions.}

\section{Introduction}
\label{sec0}
In this article motivated by the work of \cite{NAR1}, \cite{DEG}, \cite{DG1} a number of questions arise in the context of non-commutative quantum mechanics. Which conditions extend properties of the quantum mechanical covariance matrix for the conventional Heisenberg-Weyl (H.W.) algebra to the non-commutative one in arbitrary phase-space dimensions? What criteria the Wigner quasi-probability function should meet so that the transformation of covariance matrices for the two algebras be legitimate? Do exist bounds for the linear symplectic capacity of the covariance ellipsoid and its dual in this case? Can we establish any inequalities for the linear symplectic capacities of the corresponding non-commutative phase-spaces?
\par In the literature \cite{DGP}, \cite{ND} some of the above issues have been addressed and partially studied following different paths. A brief outlook of the present work which exhibits our contribution is as follows. In \textit{Sec. 2} we define the extended Heisenberg-Weyl algebra of our interest and state the transformation matrix $M$ which connects it with the familiar H.W. algebra. We focus our attention on a particular one-parameter class of transformations and discover that the skew-symmetric matrices $\theta_{ij}=\theta E_{ij}, \eta_{ij}=\eta E'_{ij}$ of the deformed algebra, should be of even dimension if one is willing to maintain  their symplectic or anti-symplectic structure. Restricting further to orthogonal Darboux transformations, the one parameter space is forced to obtain only the two values $\pm 1$, and therefore $M\in SO(4l), \, l\in\mathbb{N}$. Such kind of transformations will be used in our example exposed at the end of the manuscript. 
\par In \textit{Sect. 3} we define the quantum mechanical covariance matrix with density operator constructed for a pure state. In order to establish the connection of the covariance matrices between the two algebras, we require the Wigner quasi-probability measure to transform as in the classical probability theory. In the remainder of this section we prove (Proposition \rf{sec2 : prop1}) that if the covariance matrix of the extended algebra is positive semi-definite then the matrix $\Sigma_{W_{\Omega}}$, with elements the traces of the density operator, is positive definite, independently of the symplectic or anti-symplectic behaviour of the matrix $S(\theta,\eta)$. This result, according to our knowledge, is new and was first appeared for the standard symplectic case in \cite{NAR1}. Also, the aforementioned property will assist in proving Proposition \rf{sec3 : prop1} of the next section.
\par In \textit{Sect. 4} the third and fourth questions admit a positive answer if one diagonalizes $\Sigma_{W_{\Omega}}$ by an $\omega$-symplectic matrix (Proposition \rf{sec3 : prop1}). The proof is based on an extension of Williamson's diagonalization theorem and applied to four phase-space dimensions. The reason for this dimensional limitation is due to the technical complexity associated with the use of Sylvester's criterion for positive semi-definite matrices. Depending on whether the matrix $S$ is symplectic or anti-symplectic we derive inclusions between the Wigner ellipsoids and its duals. These inclusions, by the monotonicity property of the symplectic capacity, imply certain inequalities among them. The results of this section are also new.
\par Finally, in \textit{Sec. 5} we present an application for a simple model, borrowed from our previous work on non-commutative harmonic oscillator \cite{HS}, and justify our theoretical predictions for the ground state of the system.   
\par \textbf{Conventions:} We conclude introduction with a statement of our notations used in the manuscript. Latin indices take values in the set $\{1,2,\cdots,n\}$ whereas Greek indices are phase-space indices with values $\{1,2,\cdots,2n\}$. The superscript ``${}^{\intercal}$" indicates transposition and the over bar denotes complex conjugation. We denote by $\mathcal{M}(2n,\mathbb{R})$ the set of all $2n\times 2n$ real constant matrices and use the compact notations:
\beqr
S^{2n}=\{M \in \mathcal{M}(2n,\mathbb{R}): \, M=M^{\intercal}\}
\label{sec0 : eq1}  
\feqr
for the real symmetric matrices,
\beqr
S^{2n}_{++}=\{M \in S^{2n}: \, M\succ 0\}
\label{sec0 : eq2}  
\feqr
for the positive definite symmetric matrices,
\beqr
S^{2n}_{+}=\{M \in S^{2n}: \, M\succeq 0\}
\label{sec0 : eq3}  
\feqr 
for the semi-definite positive symmetric matrices,  
\beqr
Sp_{\sigma}(2n,\mathbb{R})=\{M \in \mathcal{M}(2n,\mathbb{R}): \sigma(Mu,Mv)=\sigma(u,v) \}
\label{sec0 : eq4}  
\feqr
for the standard symplectic group with form $\sigma(u,v)=u^{\intercal}J v$ on $\mathbb{R}^n\oplus \mathbb{R}^n$ (see \rf{sec1 : eq: 2} for the definition of J) and  
\beqr
Sp_{\omega}(2n,\mathbb{R})=\{M \in \mathcal{M}(2n,\mathbb{R}): \, \omega(Mu,Mv)=\omega(u,v) \}
\label{sec0 : eq5}  
\feqr
for the $\omega$-symplectic group corresponding to the linear transformations which leave the skew-symmetric bilinear form $\omega(u,v)=u^{\intercal}\Omega v$ on $\mathbb{R}^n\oplus \mathbb{R}^n$ invariant ($\Omega$ is defined in \rf{sec1 : eq: 6}).    
\section{A class of extended Heisenberg-Weyl algebras}
\label{sec1}
Consider the extended Heisenberg-Weyl algebra in $n\geq2$, 
\beqr
[\hat{q}_i,\hat{q}_j]=i\theta_{ij}\hat{I}, \quad [\hat{p}_i,\hat{p}_j]=i\eta_{ij}\hat{I}, \quad [\hat{q}_i,\hat{p}_j]=if(\hbar,\theta,\eta)\delta_{ij} \hat{I}, \,\, i,j=1,\cdots,n
\label{sec1 : eq: 0}
\feqr
where $f(\hbar,\theta,\eta)$ is a function of Planck's constant $\hbar=h/2\pi=6.62607015\times 10^{-34} J\cdot Hz^{-1}$ as well as the deformation parameters $\theta_{ij}, \eta_{ij}$ which are assumed to form $n\times n$ real, skew-symmetric and constant matrices. We concentrate on the case in which $\theta_{ij}=\theta E_{ij}, \,\eta_{ij}=\eta E'_{ij}$\footnote{In $n=2$ the square matrices can be written as $\theta_{ij}=\theta \epsilon_{ij},\, \eta_{ij}=\eta \varepsilon_{ij}, \, i,j=1,2$ where $\varepsilon_{ij}$ is the rank-2 Levi-Civita tensor.} where $E_{ij}, E'_{ij}=\pm1$ for $i\neq j$ and zero otherwise. The natural units of both $\hat{q}$'s and $\hat{p}$'s are to be those of length and momentum respectively. Adopting the notation $\hat{z}_{\alpha}=\hat{q}_{\alpha}, \, \alpha=1,\cdots,n$ and $\hat{z}_{\alpha}=\hat{p}_{\alpha-n}, \, \alpha=n+1,\cdots,2n$ the commutation relations can be written compactly in phase-space as
\beqr
[\hat{z}_{\alpha},\hat{z}_{\beta}]=i\left(f(\hbar,\theta,\eta)J+S\right)_{\alpha \beta}\hat{I}
\label{sec1 : eq: 1}
\feqr
where $J$ is the standard symplectic matrix
\beqr
J=
\begin{pmatrix}
	0_{n\times n} & I_{n\times n}\\
	-I_{n\times n} & 0_{n\times n}
\end{pmatrix}
\label{sec1 : eq: 2}
\feqr
and S is given by
\beqr
S(\theta,\eta)=
\begin{pmatrix} \theta_{n\times n}& 0_{n\times n}\\
	0_{n\times n} & \eta_{n\times n}. \end{pmatrix}
\label{sec1 : eq: 3}
\feqr
This algebra is related to the familiar H.W. algebra
\beqr
[\hat{Z}_{\alpha},\hat{Z}_{\beta}]=ig(\hbar) J_{\alpha \beta}\hat{I}
\label{sec1 : eq: 4}
\feqr 
through a non-canonical linear transformation, which is a diffeomorphism of phase space whose Jacobian matrix is not symplectic at every point. The map
\beqr
M\!\!: \,\,\hat{Z}\rightarrow \hat{z} \quad \textrm{where} \quad \hat{z}_{\alpha}=M_{\alpha \beta}\hat{Z}_{\beta},\quad M\in GL(2n,\mathbb{R})
\label{sec1 : eq: 5}
\feqr
is called a Darboux map and induces the matrix equation between the two algebras
\beqr
g(\hbar) MJM^{\intercal}=f(\hbar,\theta,\eta)J+S(\theta,\eta)=\Omega(\hbar,\theta,\eta).
\label{sec1 : eq: 6}
\feqr
Relation \rf{sec1 : eq: 6} implies that $M$ is not manifestly a symplectic matrix. 
If one writes $M$ in block matrix form as
\beqr
M=\begin{pmatrix}
	A_{n\times n} & B_{n\times n}\\
	C_{n\times n} & D_{n\times n}
\end{pmatrix}
\label{sec1 : eq: 7}
\feqr
then \rf{sec1 : eq: 6} leads to the following matrix system 
\beqr
g(\hbar)(AD^{\intercal}-BC^{\intercal})&=&f(\hbar,\theta,\eta)I_{n\times n} \non \\
g(\hbar)(AB^{\intercal}-BA^{\intercal})&=&\theta_{n\times n} \non \\
g(\hbar)(CD^{\intercal}-DC^{\intercal})&=&\eta_{n\times n} 
\label{sec1 : eq: 8}
\feqr
with unknowns the submatrices $A,B,C,D$. Using the possibility $AB^{\intercal},\, CD^{\intercal} $ to be both skew-symmetric\footnote{A different case, in which the identity matrix is replaced by a diagonal matrix with different elements along the main diagonal, was studied in \cite{HS}.}, we obtain the equivalent system
\beqr
g(\hbar)(AD^{\intercal}-BC^{\intercal})&=&f(\hbar,\theta,\eta)I_{n\times n} \non \\
2g(\hbar) AB^{\intercal}&=&\theta_{n\times n} \non \\
2g(\hbar)CD^{\intercal}&=&\eta_{n\times n} 
\label{sec1 : eq: 9}
\feqr
Tracing the first equation of \rf{sec1 : eq: 9} yields $f(\hbar,\theta,\eta)=g(\hbar)\textrm{Tr}(AD^{\intercal}-BC^{\intercal})/n$. The unique decomposition of a matrix into symmetric and skew-symmetric parts allows us to search for solutions of the particular form $N_{n\times n}=\lambda I_{n\times n}+\mu E_{n\times n}, \, \lambda,\mu \in \mathbb{R}$. One can verify that a viable set of solutions is
\beqr
&&A=a I_{n\times n}, \,\,\, B=b E^{\intercal}_{n\times n}, \,\, C=c E'_{n\times n}, \,\, D=d I_{n\times n}, \,\, \textrm{where} \non \\ &&ab=\frac{\theta}{2g(\hbar)},\,\, cd=\frac{\eta}{2g(\hbar)}, \,\, f(\hbar,\theta,\eta)=adg(\hbar) \left(1-\frac{\theta \eta}{4g(\hbar)^2 a^2 d^2}\right)
\label{sec1 : eq: 10}
\feqr
Special attention should be paid to the derivation of $f(\hbar,\theta,\eta)$ since the term $E'E$ equals the identity matrix on condition that $E$ is nonsingular and therefore invertible. This is true for even dimensional matrices $n=2k$ because then an antisymmetric matrix satisfies the relation $\det E_{2k\times 2k}=(\textrm{Pf} \, E_{2k\times 2k})^2>0$ where $\textrm{Pf}\, E=\varepsilon_{i_1j_1 i_2j_2 \cdots i_kj_k}E_{i_1j_1}E_{i_2j_2}\cdots E_{i_kj_k}/2^k k!$ is the Pfaffian of $E$ \cite{TV}. For square matrices with $n=2k+1$ the determinant vanishes since there are null eigenvalues. Also every $2k\times 2k$ antisymmetric matrix can be diagonalized by a unitary matrix and it has purely imaginary eigenvalues coming in conjugate pairs. Denoting these by $\pm i\lambda_j, \, j=1,\cdots,k$, with $\lambda_j$ being positive reals, $A$ can be brought into a unique canonical form, by a special orthogonal transformation $\mathcal{O}$
\beqr
\mathcal{O}^{-1}A\mathcal{O}=A_d=\bigoplus_{j=1}^k A_j, \quad A_j=\begin{pmatrix}
	0 & \pm \lambda_j\\
\mp \lambda_j & 0
\end{pmatrix}
\label{sec1 : eq: 11}
\feqr
where $\oplus_{j=1}^k A_j$ is the direct sum of skew-symmetric $2\times 2$ sub-blocks $A_j$. If in addition $A$ is an orthogonal matrix then its eigenvalues should have unit modulus and $\lambda_j=1, \forall j=1,\cdots,k$. Hence $E$ equipped with antisymmetry and orthogonality can be written in block diagonal form as  
\beqr
E_d=\bigoplus_{j=1}^k A_j, \quad A_j=\begin{pmatrix}
	0 & \pm 1\\
	\mp 1 & 0
\end{pmatrix}.
\label{sec1 : eq: 12}
\feqr
Requiring $\lim_{\theta,\eta\rightarrow 0}f(\hbar,\theta,\eta)=\hbar$, the real parameters $a,d,b,c,f(\hbar,\theta,\eta)$ should satisfy the relations  
\beqr
d=\frac{1}{a}, \,\, b=\frac{\theta}{2g(\hbar) a}, \,\, c=\frac{\eta a}{2g(\hbar)}, \, \, f(\hbar,\theta,\eta)=g(\hbar) \left(1-\frac{\theta \eta}{4g(\hbar)^2}\right)
\label{sec1 : eq: 13}
\feqr
Dimensional anlysis and experimental results \cite{CHK} demand the deformation parameters $\theta,\eta$ to satisfy the relation $\lvert\theta\eta \rvert\approx (\epsilon g(\hbar))^2$ where $\epsilon \in [0,1)$. \\
\textbf{Remark}\\
 Regarding the acceptable values of $k$ such that $E_d$ is symplectic or anti-symplectic, we distinguish the following two cases:
\begin{itemize} 
\item if $k=2l$ then $E$ is standard symplectic or anti-symplectic. The relation $E^{\intercal}JE=\pm J$, when written in blocked form, simplifies to 
\beqr
A_E^{\intercal}D_E=\pm I
\label{sec1 : eq: 14}
\feqr
which ensures that $E$ is symplectic for $A_E=D_E$ or anti-symplectic for $D_E=A_E^{\intercal}$. Notice that every anti-symplectic matrix can be factorized as the product $TS$ where $S\in Sp_{\sigma}(2n,\mathbb{R})$ and 
\beqr
T=\begin{pmatrix} I_{n\times n} & 0_{n\times n} \\ 0_{n\times n} & -I_{n\times n} \end{pmatrix}
\label{sec1 : eq: 15}
\feqr
\item if $k=2l+1, \, l\geq 1$ then $E$ is neither symplectic nor anti-symplectic. The reason is based on the simple observation that after calculating $EJE^{\intercal}$ there always exists a term of the form $A_1 \begin{pmatrix}0 & 1 \\ 0 & 0 \end{pmatrix}A_{\frac{k-1}{2}}\neq \begin{pmatrix}0 & 1 \\ 0 & 0 \end{pmatrix}$ in the center, at the top of the matrix which destroys the symplectic or anti-symplectic property of $E$.  
\end{itemize}

If $a=d=\pm 1$ and rescaling $M$ by its determinant $\det M=\sqrt{1-\frac{\theta \eta}{4g^2(\hbar)}}$, the Darboux matrix $M\in SO(4l),\, l\in \mathbb{N}$ . As a consequence, the Euclidean distance on the phase-space remains rotationally invariant. Finally, observe that if $S\in SO(2n)$ and $[S,J]=0$ then $S\in Sp(2n,\mathbb{R})$ whereas if the commutator is replaced by $\{S,J\}=0$ then $S$ is anti-symplectic. 

\section{Covariance matrix and the Robertson-Schr\"{o}dinger uncertainty principle}
\label{sec2}

We start this section by giving the definition of the covariance matrix in quantum mechanics.
\begin{definition}
\label{sec2 : eq: 0}
The quantum mechanical covariance matrix associated with an $\Omega$-Wigner quasi-probability measure $W_{\Omega}[\psi]$ for a pure state $\psi\in\mathcal{S}(\mathbb{R}^{n})$ \footnote{$\mathcal{S}(\cdot)$ is the Schwartz space of test functions whereas $\mathcal{S}'(\cdot)$ its dual.} and density operator $\hat{\rho}=\lvert \psi \rangle \langle\psi\rvert$, is defined by
\beqr
&&\!\!\!\!\!\!\!\!\!\!\!\textrm{Cov}_{W_{\Omega}}(\hat{z},\hat{z}):=\mathbb{E}_{W_{\Omega}}\left((\hat{z}-\mathbb{E}_{W_{\Omega}}(\hat{z}))(\hat{z}-\mathbb{E}_{W_{\Omega}}(\hat{z})^{\intercal})\right)=\Sigma_{W_{\Omega}}+\frac{i}{2}\Omega, \quad where \non \\ 
&&\!\!\!\!\!\!\!\!\!\!\! \left(\Sigma_{W_{\Omega}}\right)_{\alpha \beta}=\left\{\begin{array}{cc} \frac{1}{2}\textrm{Tr}(\hat{\rho}\{\hat{z}_{\alpha}, \hat{z}_{\beta}\})-\textrm{Tr}(\hat{\rho}{\hat{z}_{\alpha}})\textrm{Tr}(\hat{\rho}{\hat{z}_{\beta}}) & \alpha\neq \beta\\
	\textrm{Tr}(\hat{\rho} \hat{z}_{\alpha}^2)-(\textrm{Tr}(\hat{\rho}{\hat{z}_{\alpha}}))^2& \alpha=\beta \end{array}\right.
\label{sec2 : eq: 1}
\feqr 
where $\hat{z}_a, \, a=1,\cdots,2n$ is a row vector operator with entries the Cartesian coordinates of position and momentum operators.
\end{definition}
Recall that the expectation value of a self-adjoint operator for which $\hat{\rho}\hat{A}$ is of trace-class, is given by \cite{WO}
\beqr
\mathbb{E}_{W_{\Omega}}(\hat{A})=Tr(\hat{\rho}\hat{A})=\int_{\mathbb{R}^{2n}}W_{\Omega}[\psi](z)\sigma_{\Omega}(z)dz
\label{sec2 : eq: 2}
\feqr
where $\sigma_{\Omega}(z)$ is the Weyl symbol (or transform) of the operator $\hat{A}$. In configuration space it is obtained by
\beqr
\sigma_{\Omega}(x,p)=\frac{1}{(2\pi f(\hbar,\theta,\eta))^n}\int_{\mathbb{R}^{n}}e^{-i<\textbf{y},\textbf{p}>/f(\hbar,\theta,\eta)} <\textbf{q}-\textbf{y}/2 \lvert\hat{A}\rvert\textbf{q}+\textbf{y}/2> d\textbf{y}
\label{sec2 : eq: 3}
\feqr
with $<\textbf{q}-\textbf{y}/2 \lvert\hat{A} \rvert\textbf{q}+\textbf{y}/2>\in \mathcal{S}'(\mathbb{R}^n\times \mathbb{R}^n)$ be the kernel of the operator. Setting $\hat{A}=\hat{\rho}$ in \rf{sec2 : eq: 3} we obtain the Wigner function
\beqr  && W_{\Omega}[\psi](z)=\frac{1}{(2\pi f(\hbar,\theta,\eta))^n}\int_{\mathbb{R}^{n}}e^{-i<\textbf{y},\textbf{p}>/f(\hbar,\theta,\eta)}\psi(\textbf{q}-\textbf{y}/2)\bar{\psi}(\textbf{q}+\textbf{y}/2)d\textbf{y} \non \\
&& \textrm{where} \,\,\textbf{q},\textbf{y},\textbf{p}\in\mathbb{R}^n  
\label{sec2 : eq: 4}
\feqr
\textbf{Remarks}\\
\begin{itemize}
\item The condition which guarantees finiteness of the covariance matrix or equivalently existence up to the second moments, is
\beqr
(1+\lVert z\rVert^2)W_{\Omega}[\psi](z) \in L^1(\mathbb{R}^{2n})
\label{sec2 : eq: 5}
\feqr
where $\lVert z\rVert$ is the Euclidean norm on $\mathbb{R}^{2n}$. We also assume that 
\beqr
\int_{\mathbb{R}^{2n}} W_{\Omega}[\psi](z) dz=1.
\label{sec2 : eq: 6}
\feqr
The Wigner function is real and supports negative values. The later claim is showed by the following bound
\beqr
&& \lvert W_{\Omega}[\psi](z)\rvert\leq \left(\frac{2}{(\pi f(\hbar,\theta,\eta))}\right)^n \lVert \Phi_1\rVert_{L^2}\lVert \Phi_2\rVert_{L^2} \quad  \textrm{where}  \quad  \Phi_1(\textbf{u})=\psi(\textbf{u}) \non \\ 
&& \textrm{and} \quad \Phi_2(\textbf{u})=\bar{\psi}(\textbf{u}-\textbf{y}) e^{-i<\textbf{u},\textbf{p}>/f(\hbar)}
\label{sec2 : eq: 7}
\feqr
where a change of variables and use of the Cauchy-Schwarz inequality have been performed to \rf{sec2 : eq: 4}. From the definition \rf{sec2 : eq: 4} of Wigner function it is evident that all even wave functions reach the upper bound of \rf{sec2 : eq: 7} at $(\textbf{q},\textbf{p})=(\textbf{0},\textbf{0})$, and all odd ones the lower bound.   
\item The covariance matrix can be casted into the form
\beqr
&& \textrm{Cov}_{W_{\Omega}}(\hat{z},\hat{z})=  M\textrm{Cov}_{W_{J}}(\hat{Z},\hat{Z})M^{\intercal} \non \\
&& \textrm{where} \quad \textrm{Cov}_{W_{J}}(\hat{Z},\hat{Z})=\Sigma_{W_{J}}(\hat{Z},\hat{Z})+\frac{ig(\hbar)}{2}J. 
\label{sec2 : eq: 8}
\feqr
This expression is meaningful if we require the Wigner function to behave as a classical distribution function, therefore obey the transformation rule
\beqr
W_{\Omega}[\psi](z)=W_J[\Psi](M^{-1}z).
\label{sec2 : eq: 9}
\feqr
\item The main diagonal entries of the symmetric $\Sigma_{W_{\Omega}}$ matrix are identified to be the variances $Var_{W_{\Omega}} \hat{A}$. Definition \rf{sec2 : eq: 0} can be extended to incorporate mixed states with density operator $\hat{\rho}=\sum_{k=1}^{N} p_k \lvert\psi_k \rangle \langle\psi_k \rvert$ where  $p_k\in [0,1]$ being the weights of the states which sum up to unity, $\sum_{k=1}^N p_k=1 $. 
In the case of H.W. algebra the matrix $\Sigma_{W_J}+\frac{ig(\hbar)}{2}J$ is a non-negative Hermitian matrix \cite{NAR1} and as a consequence its determinant satisfies
\beqr
&& \!\!\!\!\!\!\!\!\!\!\!\!\det\left(\Sigma_{W_J}+\frac{ig(\hbar)}{2}J\right)\geq 0, \quad \textrm{with}\non \\ &&\!\!\!\!\!\!\!\!\!\!\!\! W_J[\Psi](Z)\!\!=\!\!\frac{1}{(2\pi g(\hbar))^n}\int_{\mathbb{R}^n}\!\!\!\! \Psi\left(\textbf{Q}-\frac{\textbf{Y}}{2}\right) \bar{\Psi} \left(\textbf{Q}+\frac{\textbf{Y}}{2}\right)e^{\frac{i}{g(\hbar)}<\textbf{Y},\textbf{P}>}d \textbf{Y}.
\label{sec2 : eq: 10}
\feqr
In two phase-space dimensions this inequality provides the well-known Robertson-Schr\"{o}dinger uncertainty principle ($R.S.U.P.$) 
\beqr
(Var_{W_{J}} \hat{Q})(Var_{W_{J}} \hat{P})-\left( \frac{1}{2}\textrm{Tr}(\hat{\rho}\{\hat{Q},\hat{P}\})-\textrm{Tr}(\hat{\rho}\hat{Q})\textrm{Tr}(\hat{\rho}\hat{P}) \right)^2 \geq\frac{g(\hbar)^2}{4}
\label{sec2 : eq: 11}
\feqr
which is the strong version of the Heisenberg uncertainty principle. 
\end{itemize}
\par We now prove a technical but useful proposition which extends Lemma (2.3) of \cite{NAR1} to the non-commutative phase-space. 
\begin{proposition}
\label{sec2 : prop1}
Let $\Sigma_{W_{\Omega}}\in S^{2n}$ with $\hbar, \theta, \eta$ real numbers. Then the following hold:
\begin{enumerate}
\item [$i)$] The matrix $\Sigma_{W_{\Omega}}(\hat{z},\hat{z})+\frac{i}{2}\Omega(\hbar,\theta,\eta)$ has real eigenvalues.
\item [$ii)$] If there exist real numbers $\hbar,\theta,\eta \neq 0$ such that $S$ is symplectic or anti-symplectic, and the covariance matrix is positive semi-definite then $\Sigma_{W_{\Omega}}\in S^{2n}_{++}$ and $Cov_{W_{\Omega}}(\hat{z},\hat{z};\hbar',\theta')\succeq 0, \, \forall \, \hbar'\leq \hbar, \theta'\leq \theta$ as long as $f(\hbar,\theta)$ is an homogeneous function of degree one. 	
\end{enumerate}
\end{proposition}
\textbf{Proof}
\begin{enumerate}
\item[$i)$] Using  that $\Sigma_{W_{\Omega}}\in S^{2n}$ as well as the identities $(iJ)^\dagger=-iJ^{\intercal}=iJ$ and $(iS)^{\dagger}=iS$ the hermiticity of the covariance matrix is an immediate consequence.
\item[$ii)$] Suppose now $\Sigma_{W_{\Omega}}$ has a negative eigenvalue $\lambda$ corresponding to a real eigenvector $v_{\lambda}$. Since $J, \, S(\theta,\eta)$ are skew-symmetric matrices, $v_{\lambda}^{\intercal}J v_{\lambda}=0=v_{\lambda}^{\intercal}S(\theta,\eta) v_{\lambda}$ and the Rayleigh quotient is negative 
\beqr
R(Cov_{W_{\Omega}},v_{\lambda})=\frac{\langle v_{\lambda},Cov_{W_{\Omega}} v_{\lambda}\rangle}{\lVert v_{\lambda}\rVert^2}=\lambda <0
\label{sec2 : eq: 12}
\feqr
This contradicts the assumption $Cov_{W_{\Omega}}(\hat{z},\hat{z};\hbar,\theta,\eta)\succeq 0$ and therefore $\Sigma_{W_{\Omega}}$ is a non-negative matrix. 
\par To complete the proof we now show that $\Sigma_{W_{\Omega}}$ cannot have a zero eigenvalue. Let $v_0$ be a real eigenvector of $\Sigma_{W_{\Omega}}$ with zero eigenvalue and consider the perturbed complex vector 
\beqr
v_{\lambda}(\epsilon_{\hbar}, \epsilon_{\theta},\epsilon_{\eta})&=&\left(I+i(f(\epsilon_{\hbar},\epsilon_{\theta},\epsilon_{\eta})J+S(\epsilon_{\hbar},\epsilon_{\theta},\epsilon_{\eta}))\right)v_0 \non \\
&=&\left(I+i(f(\epsilon)J+S(\epsilon)\right)v_0
\label{sec2 : eq: 13}
\feqr 
where in brief we denote by $f(\epsilon)=f(\epsilon_{\hbar},\epsilon_{\theta},\epsilon_{\eta})$ and by $S(\epsilon)=S(\epsilon_{\hbar},\epsilon_{\theta},\epsilon_{\eta})$. Performing a lengthy but straightforward calculation and taking into account:
\begin{itemize}
\item  $\theta_{ij}=-\eta_{ij}$ ($S$ is anti-symplectic)
\item  the identities: $\Sigma_{W_{\Omega}} v_0=v_0^{\intercal}\Sigma_{W_{\Omega}}=0$, $v_0^{\intercal}Jv_0=0=v_0^{\intercal}S v_0$, $\lVert Jv_0 \rVert^2=\lVert v_0 \rVert^2$, $v_0^{\intercal}S(\theta,\eta) S(\epsilon)v_0 =-\theta \epsilon_{\theta}\lVert v_0 \rVert^2$ and
\item  disregarding terms but linear in $\epsilon$'s
\end{itemize}
we end up with the expression
\beqr
\bar{v}_{\lambda}^{\intercal}Cov_{W_{\Omega}}(\hat{z},\hat{z};\hbar,\theta) v_{\lambda}=(f(\epsilon_{\hbar},\epsilon_{\theta})f(\hbar,\theta)+\epsilon_{\theta}\theta) \lVert v_0 \rVert^2
\label{sec2 : eq: 14}
\feqr
For both $\lvert \epsilon_{\hbar} \rvert,\lvert\epsilon_{\theta}\rvert$ small enough and $f(\epsilon_{\hbar},\epsilon_{\theta})f(\hbar,\theta),\theta\epsilon_{\theta}<0$ we get a negative result which again contradicts the hypothesis of positive semi-definiteness. If we relax the condition of $S$ being anti-symplectic then it appears the extra contribution
\beqr
-\frac{1}{2}v_0^{\intercal}(f(\epsilon)\{S(\theta,\eta),J\}+f(\hbar,\theta)\{J,S(\epsilon)\})v_0=-(f(\epsilon)\theta+f(\hbar,\theta)\epsilon_{\theta})v_0^{\intercal}SJv_0 \non \\
\label{sec2 : eq: 15}
\feqr
where $SJ$ is symmetric with a minimum eigenvalue $\lambda_{min.}(SJ)=-1<0$. Therefore the above expression is again negative for $f(\epsilon)\theta<0, \, f(\hbar,\theta)\epsilon_{\theta}<0$ and leads to a contradiction.
\par By hypothesis, $f(s\hbar,s\theta)=sf(\hbar,\theta)$, $s\in(0,1]$, and
\beqr
Cov_{W_{\Omega}}(\hat{z},\hat{z};\hbar',\theta')&=&\Sigma_{W_{\Omega}}+\frac{i}{2}s(f(\hbar,\theta)J+\theta S) \non \\
&=&(1-s)\Sigma_{W_{\Omega}}+s(\Sigma_{W_{\Omega}}+\frac{i}{2}(f(\hbar,\theta)J+\theta S)\succeq 0.
\label{sec2 : eq: 16}
\feqr
Relation \rf{sec2 : eq: 16} is true since $(1-s)\Sigma_{W_{\Omega}}\succeq 0$ and $Cov_{W_{\Omega}}(\hat{z},\hat{z};\hbar,\theta)\succeq 0$.
\end{enumerate}
Another useful observation, which clarifies the interplay of semi-definite positivity for the covariance matrices of the corresponding algebras, is stated by the following proposition.
\begin{proposition}
\label{sec2 : prop2}
Let $\mathcal{P}_{\omega}\in Sp_{\omega}(2n,\mathbb{R})$ and $\mathcal{P}_{\omega}=MSM^{\intercal}$ where the Darboux transformation $M\in SO(4l)$ satisfies $g(\hbar) MJM^{\intercal}=\Omega$ and $S\in Sp_{\sigma}(2n,\mathbb{R})$. The matrix $\mathcal{P}_{\omega}^{\intercal}\Sigma_{W_{\Omega}}(\hat{z},\hat{z})\mathcal{P}_{\omega}+\frac{i}{2}\Omega$ is positive semi-definite iff $\Sigma_{W_{J}}(\hat{Z},\hat{Z})+\frac{ig(\hbar)}{2}J$ is positive semi-definite.
\end{proposition}
\textbf{Proof}\\
Theorem (2.2) of \cite{NAR1} guarantees $\Sigma_{W_{J}}(\hat{Z},\hat{Z})+\frac{ig(\hbar)}{2}J\succeq 0$ and Williamson's standard symplectic theorem implies the existence of $S\in Sp_{\sigma}(2n,\mathbb{R})$ which diagonalizes $\Sigma_{W_J}$. Therefore we have the following relations
\beqr
S^{\intercal}\left(\Sigma_{W_J}+\frac{i}{2}g(\hbar)J\right) S\succeq 0 &\Longleftrightarrow& M\left(S^{\intercal}\Sigma_{W_J}S+\frac{i}{2}g(\hbar)J\right)M^{\intercal} \succeq 0 \non \\
&\Longleftrightarrow&(MS^{\intercal}M^{\intercal})\left(\Sigma_{W_{\Omega}}+\frac{i}{2}\Omega \right)(MSM^{\intercal}) \succeq 0 \non \\
&\Longleftrightarrow&\mathcal{P}_{\omega}^{\intercal}\Sigma_{W_{\Omega}}\mathcal{P}_{\omega}+\frac{i}{2}\Omega \succeq 0
\label{sec2 : eq: 17}
\feqr 
which prove the claim. A point that requires attention is the particular $\mathcal{P}_{\omega}$ which does not diagonalize $\Sigma_{W_{\Omega}}$ as one might suspect. This situation will be partly encountered in our toy model since there $\Sigma_{W_{J}}$ will be diagonal by construction.    
\par The decomposition of an $\omega$-symplectic matrix into a Darboux and a standard symplectic matrix was also pointed out by \cite{ND}. In our case $M\in SO(4l)$ as explained in section 2. Also $\mathcal{P}_{\omega}$ turns out to be a special orthogonal matrix as well, since $\mathcal{P}_{\omega}\mathcal{P}_{\omega}^{\intercal}=I$ and $\det \mathcal{P}_{\omega}=(\det M)^2 \det S=1$. Recall that a standard symplectic matrix always has unit determinant.
\par The determinant and trace of the covariance matrices for the corresponding algebras are related through the following expressions
\beqr
\det \textrm{Cov}_{W_{\Omega}}(\hat{z},\hat{z})&=& \det \textrm{Cov}_{W_{J}}(\hat{Z},\hat{Z})\geq 0
\label{sec2 : eq: 18} \\
Tr(Cov_{W_{\Omega}}(\hat{z},\hat{z};\hbar,\theta))&=&Tr(Cov_{W_{J}}(\hat{Z},\hat{Z};\hbar,\theta))\geq 0.
\label{sec2 : eq: 19}
\feqr 
In the $\theta\rightarrow 0$ limit and slightly modifying the notation, namely $\hat{Z}^{\intercal}=(\hat{Q}_1,\hat{P}_1;\cdots;\hat{Q}_n,\hat{P}_n)^{\intercal}$, the covariance matrix is easily recognizable.
\section{Symplectic capacity and R.S.U.P.}
\label{sec3}
In paper \cite{DEG1} the R.S.U.P., in two phase-space dimensions ($n=1$), was expressed in terms of the area of the covariance ellipse. It was then proved that $Area(\mathcal{B})=2\pi \sqrt{\det(Cov_{W_J}(\hat{Z},\hat{Z}))}\geq h/2 $. The same idea can be extended in higher dimensions but again in terms of areas of the intersections of the conjugate planes $q_j,p_j$ with the covariance (or Wigner) ellipsoid. In general it is defined as follows:  
\begin{definition}
\label{sec3 : def1}
The Wigner ellipsoid associated to $\Sigma_{W_{\Omega}}$ is defined by the quadratic form
\beqr
\mathcal{E}_{\Sigma_{W_{\Omega}}}\equiv\{z\in\mathbb{R}^{2n}: \, \, \frac{1}{2}\langle z,\Sigma_{W_{\Omega}}^{-1}z \rangle\leq 1\}
\label{sec3 : eq1}
\feqr
and its Legendre transorm (or dual) in phase-space, is defined by
\beqr
\mathcal{E}_{\Sigma_{W_{\Omega}}}^*\equiv\{\zeta\in\mathbb{R}^{2n}: \, \, \frac{1}{2}\langle \zeta,\Sigma_{W_{\Omega}}\zeta \rangle \leq 1\}
\label{sec3 : eq2}
\feqr
\end{definition}
\par We now introduce the notion of symplectic capacity \cite{DG1}, \cite{MS} and study its connection to R.S.U.P. 
\begin{definition}For a symplectic manifold $(\mathbb{R}^n\oplus \mathbb{R}^n,\sigma)$ of fixed dimension $2n$ it is defined through the map
\beqr
(\mathcal{A},\sigma) \rightarrow c(\mathcal{A})\in [0, +\infty], \, \, \forall \mathcal{A}\subset \mathbb{R}^n\oplus \mathbb{R}^n
\label{sec3 : eq3}
\feqr
which satisfies the following properties:
\begin{enumerate} 
\item[$(3.1)$] Monotonicity: If there exists a symplectic embedding: $\phi: \,\,  \mathcal{A}_1\rightarrow \mathbb{R}^{2n}$ such that $\phi(\mathcal{A}_1)\subset \mathcal{A}_2$ then $c(\mathcal{A}_1)\leq c(\mathcal{A}_2)$ 
\item[$(3.2)$] Conformality: $c(\lambda\mathcal{A})=\lambda^2 c(\mathcal{A}), \, \textrm{for}\,\, \lambda\in \mathbb{R}$
\item[$(3.3)$] Nontriviality: $c(\mathcal{B}^{2n}(1),\sigma_0)>0, \, \, c(\mathcal{Z}^{2n}(1)<\infty$ for the open unit ball $\mathcal{B}^{2n}(1)$ and the open symplectic cylinder $\mathcal{Z}^{2n}(1)$ in the standard space $(\mathbb{R}^{2n},\sigma_0)$.
\end{enumerate}
\end{definition}
The last property for $c(\mathcal{B}^{2n}(1),\sigma_0)=\pi=c(\mathcal{Z}^{2n}(1)$ is equivalent to Gromov's nonsqueezing theorem. In what follows we will consider the linear symplectic capacity which obeys the extra property 
\beqr
c_{lin.}(\phi(\mathcal{A}))=c_{lin.}(\mathcal{A}), \, \, \forall \phi\in ASp(2n,\mathbb{R})
\label{sec3 : eq4}
\feqr
where $ASp(2n,\mathbb{R})$ is the affine symplectic group consisting of translations accompanied by symplectic transformations on phase-space.
\par Williamson's $\omega$-symplectic diagonilization theorem plays a prominent role in proving Proposition \rf{sec3 : prop1}. Its proof follows similar steps to the standard case \cite{JW}, \cite{PSL} and states:
\theorem{ \label{sec3 : th1}
Let $A\in S^{2n}_{++}$, then there exists $\mathcal{P}_{\omega}\in Sp_{\omega}(2n,\mathbb{R})$ such that  
\beqr
\mathcal{P}_{\omega}^{\intercal}A \mathcal{P}_{\omega}=\begin{pmatrix} \Lambda_{\omega} & 0 \\ 0 & \Lambda_{\omega}\end{pmatrix}=W_{\omega}
\label{sec3 : eq8}
\feqr
where $\Lambda_{\omega}=diag(\lambda_{1,\omega},\cdots,\lambda_{n,\omega})$ and the $\lambda_{j,\omega}>0$ being the $\omega$-symplectic eigenvalues of $A$ satisfying 
\beqr
det(\Omega A\pm i \lambda_{j,\omega} I)=0, \quad j=1,\cdots,n
\label{sec3 : eq4a}
\feqr
}
\par \normalfont The diagonal matrix $W_{\omega}$ is called the ``$\omega$-Williamson form" of A. We denote by 
\beqr
Spec_{\omega}(A)=\{\lambda_{1,\omega},\cdots,\lambda_{n,\omega}\}, \quad \lambda_{1,\omega}\geq \cdots\geq \lambda_{n,\omega}>0
\label{sec3 : eq9}
\feqr
the decreasing sequence of the symplectic eigenvalues of $A$ and is called the $\omega$-symplectic spectrum of $A$. The symplectic spectrum of A coincides to its Euclidean spectrum only if $\mathcal{P}_{\omega}$ is supplemented by orthogonality.
\par We continue our study by proving the following geometric result for the non-commutative four-dimensional phase-space ($n=2$).
\begin{proposition}
\label{sec3 : prop1}
Let  $\Sigma_{W_{\Omega}} + \frac{i}{2}\Omega\succeq 0$. This condition is equivalent to the following two:
\begin{enumerate}
\item[$i)$] The linear symplectic capacity of the ellipsoid $\mathcal{E}_{\Sigma_{W_{\tilde{\Omega}}}}$, in four phase-space dimensions, is such that
\beqr
c_{lin.}(\mathcal{E}_{\Sigma_{W_{\Omega}}})\geq \pi\sqrt{f^2(\hbar,\theta)+\theta^2}
\label{sec3 : eq5}
\feqr
\item[$ii)$] The linear symplectic capacity of the dual ellipsoid $\mathcal{E}_{\Sigma_{W_{\Omega}}}^*$, in four phase-space dimensions, is such that
\beqr
c_{lin.}(\mathcal{E}_{\Sigma_{W_{\Omega}}}^*)\leq \frac{2\pi}{\sqrt{f^2(\hbar,\theta)+\theta^2}}
\label{sec3 : eq6}
\feqr
\end{enumerate}
\end{proposition}
\textbf{Proof}\\
$\Sigma_{W_{\Omega}}\in S^{2n}$ by construction, and moreover it is positive definite by Proposition \rf{sec2 : prop1}. Setting $B=\Sigma_{W_{\Omega}}^{-1}/2$ in the first condition of Proposition \rf{sec3 : prop1} and applying Williamson's theorem \rf{sec3 : th1} we get
\beqr
\frac{1}{2}{P}^{\intercal}_{\omega}B^{-1}\mathcal{P}_{\omega}+ \frac{i}{2}\Omega= \frac{1}{2}D_{\omega}^{-1} + \frac{i}{2}\Omega \succeq 0
\label{sec3 : eq10}
\feqr
where $D_{\omega}^{-1}=diag((\lambda_{1,\omega})^{-1},(\lambda_{2,\omega})^{-1};(\lambda_{1,\omega})^{-1},(\lambda_{2,\omega})^{-1})$ and the eigenvalues $\lambda_{j,\omega}^{-1}$ are of order at least $\hbar$. The characteristic polynomial is then given by
\beqr
&& det\left(\frac{1}{2}D_{\omega}^{-1} + \frac{i}{2}\Omega-\mu I\right)=
\left(\prod_{j=1}^2 F_j(\mu;\hbar,\theta)\right)+\frac{\theta^2}{16}
\left(\frac{1}{\lambda_{1,\omega}}-\frac{1}{\lambda_{2,\omega}}\right)^2=0, \,\, \textrm{where}\non \\
&& F_j(\mu;\hbar,\theta)=\mu^2-\frac{\mu}{\lambda_{j,\omega}}+\frac{1}{4}\left(\frac{1}{(\lambda_{j,\omega})^2}-(f^2(\hbar,\theta)+\theta^2)\right).
\label{sec3 : eq11}
\feqr
It has degree four in $\mu$ and its coefficients appear with alternating sign. To prove this property we combine Sylvester's citerion for Hermitian, positive semi-definite matrices \cite{HJ} with the ordering of eigenvalues. The no negativity of the principal minors\footnote{There are $2^{2n} -1$ principal minors including the determinant $det\left(D_{\omega}^{-1} + i\Omega\right)$. In our case, $n=2$, there are four non-negative principal minors of size three, producing three independent inequalities.}   of size three leads to
\beqr
f^2(\hbar,\theta)+\theta^2\leq \frac{1}{\lambda_{1,\omega}^2} \leq  \frac{1}{\lambda_{1,\omega} \lambda_{2,\omega}}\leq \frac{1}{\lambda_{2,\omega}^2}
\label{sec3 : eq13}
\feqr
which proves the claim. Moreover, using Descarte's sign rule, the characteristic polynomial has four positive real roots, explicitly given by 
\beqr
&&\mu_{(\pm,\pm),\omega}(\hbar,\theta)=\frac{1}{4}\left(\frac{1}{\lambda_{1,\omega}}+\frac{1}{\lambda_{2,\omega}}\pm\sqrt{\left(\frac{1}{\lambda_{1,\omega}}-\frac{1}{\lambda_{2,\omega}}\pm 2f(\hbar,\theta)  \right)^2 +4 \theta^2}\right), \non \\
&& \lambda_{1,\omega}\neq\lambda_{2,\omega}
\label{sec3 : eq14}
\feqr  
In view of Proposition (94) of \cite{DG1}, we find that
\beqr
c_{lin.}(\mathcal{E}_{\Sigma_{W_\Omega}})&=&\frac{\pi}{\lambda_{max.,\omega}}\geq \pi\sqrt{f^2(\hbar,\theta)+\theta^2} 
\label{sec3 : eq: 15}\\
c_{lin.}(\mathcal{E}_{\Sigma_{W_\Omega}}^*)&=&2\pi\lambda_{min.,\omega} \leq \frac{2\pi}{\sqrt{f^2(\hbar,\theta)+\theta^2}}
\label{sec3 : eq: 16}
\feqr
\textbf{Remarks}\\
\begin{itemize}
\item From \rf{sec3 : eq: 15} and using  \rf{sec1 : eq: 13} we have
\beqr
c_{lin.}(\mathcal{E}_{\Sigma_{W_\Omega}})\geq \pi \sqrt{g^2(\hbar)(1-\frac{\theta \eta}{4g(\hbar)^2})^2+\theta^2}. 
\label{sec3 : eq: 17}
\feqr
One can derive the following inclusions $\mathcal{E}_{\Sigma_{W_{\Omega^S}}}\subseteq \mathcal{E}_{\Sigma_{W_J}}\subseteq \mathcal{E}_{\Sigma_{W_{\Omega^A}}}$ where $\Omega^{S(A)}$ denotes the case in which $\Omega$ contains a symplectic ($\theta=\eta$) or an anti-symplectic ($\theta=-\eta$) $S$ matrix respectively. 
\item In $n\geq 8$ phase-space dimensions the analysis is more involved but standard.
\end{itemize}

\section{Application to a toy model}  
\label{sec4}
As an example \cite{HS} consider the non-commutative, isotropic, harmonic oscillator in four phase-space dimensions satisfying the algebra ($f(\hbar)=\hbar, \, g(\hbar,\theta)=\sqrt{\hbar^2+\theta^2}$ according to the notation used in Sect. \rf{sec1}) 
\beqr
[\hat{q}_i,\hat{p}_j]=i\hbar \delta_{ij} \hat{I}, \,\, [\hat{q}_1,\hat{q}_2]=i\theta \hat{I}, \,\, [\hat{p}_1,\hat{p}_2]=-i\theta \hat{I}, \quad i,j=1,2
\label{sec4 : eq1}
\feqr
The connection of this extended algebra with the H.W. one, is given through the orthogonal Darboux matrix
\beqr
M=\begin{pmatrix}
	1 & 0 & 0 & 0\\
	0 & \frac{\hbar}{\sqrt{\hbar^2+\theta^2}} & -\frac{\theta}{\sqrt{\hbar^2+\theta^2}} & 0 \\
	0 & \frac{\theta}{\sqrt{\hbar^2+\theta^2}} & \frac{\hbar}{\sqrt{\hbar^2+\theta^2}} & 0 \\
	0 & 0 & 0 &1
\end{pmatrix} 
\label{sec4 : eq2}
\feqr
The energy eigenvalues are found to be $E_{(n_1,n_2)}= \tilde{\omega}\sqrt{\hbar^2+\theta^2} \sum_{i=1}^2 \left(n_i+1/2\right)$ and the trajectories on phase-space  with $E_{(n_1,n_2)}$ energy satisfy the equation of an ellipsoid which is rotationally invariant. The intersection of this ellipsoid by a plane of conjugate variables $q_i,p_i$ is an ellipse with area  $2\pi\sqrt{\hbar^2+\theta^2}(n_1+n_2+1), \, n_1,n_2=0,1,2,\cdots$. This area has a lower bound, corresponding to the ground state, and given by
\beqr
Area_{min.}=2\pi\sqrt{\hbar^2+\theta^2}.
\label{sec4 : eq3}
\feqr 
\par  The Wigner function for this system, in terms of the Laguerre's polynomials \cite{HS}
\beqr
&& L_{n_i}(\tilde{Z}_i)=\frac{1}{n_i!}e^{\tilde{Z}_i}\frac{d^{n_i}}{d^{n_i}\tilde{Z}_i}\left(e^{-\tilde{Z}_i}\tilde{Z}_i^{n_i}\right), \quad \tilde{Z}_i=2\left(\frac{1}{\alpha^2}Q_i^2+\frac{\alpha^2}{\hbar^2+\theta^2}P_i^2\right) \non \\
&&\textrm{where} \,\, i=1,2, \, \, \alpha^2=\frac{\sqrt{\hbar^2+\theta^2}}{m\tilde{\omega}},
\label{sec4 : eq4}
\feqr
is given by
\beqr
W_{J}[\Psi_{(n_1,n_2)}](\tilde{Z}_1,\tilde{Z}_2)=C e^{-\frac{1}{2}(\tilde{Z}_1+\tilde{Z}_2)}L_{n_1}(\tilde{Z}_1)L_{n_2}(\tilde{Z}_2)
\label{sec4 : eq5}
\feqr
where $C$ is the normalization constant determined by  
\beqr
\int_{\mathbb{R}^4}W_{J}[\Psi_{(n_1,n_2)}](Z)dZ=1.
\label{sec4 : eq6}
\feqr
Notice that the Wigner function is an even function with respect to $Q_i, P_i$ and therefore the off-diagonal elements of $\Sigma_{W_{J}}$ vanish. 
\par The ground state is given by the Gaussian density function
\beqr
\Psi_{(0,0)}(\textbf{Q})=\left(\frac{1}{\pi^{1/4} \sqrt{a}}\right)^2 e^{-\frac{1}{2 a^2} \textbf{Q}^2}
\label{sec4 : eq7}
\feqr
and the normalized Wigner function is then found to be
\beqr
W_{J}[\Psi_{(0,0)}](\textbf{Q},\textbf{P})=\frac{1}{(\pi \sqrt{\hbar^2+\theta^2})^2} e^{-\frac{1}{ a^2} \textbf{Q}^2} e^{-\frac{a^2}{\hbar^2+\theta^2} \textbf{P}^2}.
\label{sec4 : eq8}
\feqr
This result is in agreement with Hudson \cite{HSC} and Soto, Claverie theorem \cite{SC} according to which the Wigner measure of a state vector $\Psi \in L^2(\mathbb{R}^{2n})$ is non-negative iff $\Psi$ is a Gaussian state. The diagonal elements of $\Sigma_{W_{J}}$ with multiplicity two are given by 
\beqr
\lambda_{1,max.}=Var_{W_J}(Q_i)&=& \frac{\sqrt{\hbar^2+\theta^2}}{2m \tilde{\omega}}>0 \non \\
\lambda_{2,min}=Var_{W_J}(P_i)&=& \frac{m\tilde{\omega}}{2}\sqrt{\hbar^2+\theta^2}>0, \quad i=1,2
\label{sec4 : eq9}
\feqr
assuming the system to be in the ground state. If $m\tilde{\omega}=1$ then the ellipsoid degenerates to a sphere of radius $R=\sqrt{2}(\hbar^2+\theta^2)^{1/4}$ and then $\lambda_{1,max.}=\lambda_{2,min}=\sqrt{\hbar^2+\theta^2}/2$.  We can check that the R.S.U.P. in four phase-space dimensions, namely,
\beqr
\prod_{i=1}^{2} (Var Q_i Var P_i)-\frac{1}{4}(\hbar^2+\theta^2)\sum_{i=1}^2 (Var Q_i Var P_i)+\frac{(\hbar^2+\theta^2)^2}{16}\geq 0
\label{sec4 : eq10}
\feqr
is saturated by $\Psi_{(0,0)}$ and $\det(\textrm{Cov}_{W_{\Omega}}(\hat{z},\hat{z}))=0$ using \rf{sec2 : eq: 18}. 
\par Exploiting Proposition \rf{sec3 : prop1} for the ordinary H.W. algebra we find the product of the eigenvalues to be given by 
\beqr
\prod_{j=1}^2 \frac{1}{4}\left(\frac{1}{(\lambda_{j,J})^2}-(\hbar^2+\theta^2)\right)
\label{sec4 : eq11}
\feqr
and thus the linear symplectic capacity of the Wigner ellipsoid and its dual, satisfy the predicted bounds. The covariance matrix of the extended algebra using \rf{sec2 : eq: 8} is
\beqr
\Sigma_{\Omega}=\begin{pmatrix} \lambda_1 & 0 & 0 & 0\\
0 & \frac{\theta^2 \lambda_1+\hbar^2 \lambda_2}{\hbar^2+\theta^2} & \frac{(\lambda_2-\lambda_1)\hbar \theta}{\hbar^2+\theta^2} & 0 \\
0 & \frac{(\lambda_2-\lambda_1)\hbar \theta}{\hbar^2+\theta^2} & \frac{\hbar^2 \lambda_1+\theta^2 \lambda_2}{\hbar^2+\theta^2} & 0 \\
0 & 0 & 0 & \lambda_2
\end{pmatrix}
\label{sec4 : eq12}
\feqr  
We distinguish the following two cases:
\begin{itemize}
\item $m\tilde{\omega}=1$. The matrix $\Sigma_{W_\Omega}$ is diagonal and has one eigenvalue with multiplicity four. As a consequence $c_{lin.}(\mathcal{E}_{\Sigma_{W_\Omega}})=c_{lin.}(\mathcal{E}_{\Sigma_{W_J}})$.
\item  $m\tilde{\omega}\neq 1$. The $\omega$-symplectic spectrum of $\Sigma_{W_\Omega}$ is
\beqr
Spec_{\omega}(\Sigma_{W_\Omega})=\{\lambda_{\omega}, \lambda_{\omega}, \lambda_{\omega}, \lambda_{\omega}\}, \,\, \lambda_{\omega}=\sqrt{(\hbar^2+\theta^2)\lambda_1 \lambda_2}=\frac{1}{2}(\hbar^2+\theta^2)
\label{sec4 : eq13}
\feqr
where \rf{sec4 : eq9} have been used. This result is also in agreement with the bounds predicted by Proposition \rf{sec3 : prop1}. 
\end{itemize}
\textbf{Remark}\\
The ordering of the eigenvalues
\beqr
\lambda_{2,min}\leq  \lambda_{1,max}\leq \lambda_{\omega}
\label{sec4 : eq14}
\feqr
implies the inequalities for the linearized symplectic capacities
\beqr
c_{lin.}(\mathcal{E}_{\Sigma_{W_J}})\geq c_{lin.}(\mathcal{E}_{\Sigma_{W_\Omega}}), \quad 
c_{lin.}(\mathcal{E}_{\Sigma_{W_J}}^*)\leq c_{lin.}(\mathcal{E}_{\Sigma_{W_\Omega}}^*)
\label{sec4 : eq15}
\feqr
\section{Conclusion}
In the present work for a particular one-parameter family of extended H.W. algebra we study the conditions under which the Darboux matrix is symplectic or anti-symplectic. It turns out that for specific values of the $a$-parameter, it belongs to the special orthogonal group $SO(4l), \, l\in \mathbb{N}$. 
\par The covariance matrices of the extended and ordinary H.W. algebras are meaningful provided the Wigner quasi-probability measure transforms like a classical distribution function under symplectic automorphisms. In Proposition \rf{sec2 : prop1} we have proved that the symmetric part of the covariance matrix is positive definite by imposing precise algebraic conditions. 
\par The connection of the linear symplecic capacity with the R.S.U.P. is derived in Proposition \rf{sec3 : prop1} with the assistance of the Williamson's $\omega$-symplectic diagonalization Theorem \rf{sec3 : th1}. Although the investigation is restricted in four phase-space dimensions it can also be extended to higher ones. Inclusions of the Weyl ellipsoids which depend on the symplectic or anti-symplectic nature of $S(\theta,\eta)$ lead to inequalities for the corresponding linear symplecic capacities.    
\par Finally, the rigorous agreement of our theoretical results is exposed by explicitly studying a toy model inspired by our previous work on non-commutative harmonic oscillator. \\



\end{document}